\documentclass[conference]{IEEEtran}

\usepackage[T1]{fontenc}
\usepackage[utf8]{inputenc}
\usepackage{cite}
\usepackage{url}
\usepackage{array}
\usepackage{placeins}
\usepackage{tabularx}
\usepackage{adjustbox}
\usepackage{hyperref}
\usepackage{amsmath}
\usepackage{booktabs}
\usepackage{enumitem}
\usepackage{graphicx}
\usepackage{xcolor}
\usepackage{tikz}
\usetikzlibrary{positioning, shapes, arrows, patterns, fit}

\hypersetup{
    colorlinks=true,
    linkcolor=black,
    citecolor=black,
    urlcolor=blue
}

\title{Generative AI and Copyright Infringement: A Legal-Technical Analysis of AI Music Generation Systems Under 17 U.S.C. Title 17}

\author{
\IEEEauthorblockN{Zuhaib Hussain Butt}
\IEEEauthorblockA{
Gujranwala, Punjab, Pakistan \\
Email: Zuhaibbutt3@gmail.com
}
}

\begin{document}

\maketitle

\begin{abstract}
Generative artificial intelligence (GenAI) has enabled users to synthesize music with text prompts, combining copyrighted lyrics, AI-composed melodies, and synthetic vocals that imitate real artists. This paper examines the legal and technical dimensions of AI-based music creation (e.g., Google Gemini's music tools) under U.S. copyright law. We analyze whether a user who inputs one artist's protected lyrics into a GenAI system, directs it to use another artist's voice or style, publishes the resulting song, and monetizes it violates 17 U.S.C. \S 106's exclusive rights \cite{uscode106}. The analysis integrates Title 17 doctrine (rights of reproduction, derivative works, distribution), 17 U.S.C. \S 114's narrow sound recording protection \cite{uscode114}, and the new voice-cloning laws emerging at the state level \cite{elvis}. We argue that unauthorized lyric copying poses a high risk of infringement of the musical composition, whereas mere AI-generated voice imitation typically falls outside federal sound-recording protection and instead implicates state publicity rights \cite{lehrman, richardson}. Recent cases and legislation (Concord v. Anthropic \cite{concord}; Kadrey v. Meta \cite{kadrey}; Lehrman v. Lovo \cite{lehrman}; Tennessee's "ELVIS Act" \cite{elvis}; UMG v. Uncharted Labs \cite{umg}; etc.) illustrate this split. We map AI technical components (prompt encoding, latent diffusion, neural vocoders, speaker embeddings) to legal risks, and identify a regulatory gap: federal law robustly protects lyrics and melody, but currently provides limited remedies for synthesized vocal likeness \cite{rothman, mcjohn}. The paper concludes with policy suggestions for clearer rules on AI music creation.
\end{abstract}

\begin{IEEEkeywords}
Generative AI, copyright infringement, music generation, AI voice cloning, Title 17, lyrics, sound recording, derivative works, right of publicity
\end{IEEEkeywords}

\section{Executive Summary}
Generative AI music tools (such as Google Gemini's Lyria or other multi-stage pipelines) allow users to create songs by supplying prompts that include copyrighted lyrics and instructions to emulate a particular singer's voice. This raises complex questions under U.S. copyright law \cite{uscode102}. This review finds that:

\begin{itemize}
\item \textbf{Lyrics and Composition Rights:} If a user inputs copyrighted lyrics (or even large portions) into a GenAI system and generates a new song, the output likely infringes the musical composition. Lyrics are literary works protected as part of the composition (17 U.S.C. \S 102(a)(2)) \cite{uscode102}, so copying them without permission or license violates the composer's exclusive reproduction and derivative-work rights \cite{uscode106}. Even if the AI creates a new melody, unauthorized lyric reproduction is prima facie infringement absent fair use \cite{uscode107}.

\item \textbf{Voice Cloning vs. Sound Recordings:} AI-generated "sound-alike" vocals typically do not infringe sound recordings unless actual fixed samples were copied. Section 114(b) \cite{uscode114} explicitly permits independent re-recordings or imitations of a vocal style. The recent decision in \textit{Richardson v. Kharbouch} (7th Cir. 2025) held that sounding like French Montana without sampling his track was not infringement \cite{richardson}. Likewise, \textit{Lehrman v. Lovo} (S.D.N.Y. 2025) found no federal infringement when an AI cloned professional voices for training text-to-speech \cite{lehrman}.

\item \textbf{Publicity and State Law:} Because federal copyright often fails to cover AI voice cloning, injured performers must rely on state rights of publicity or unfair competition \cite{midler, waits}. For example, Tennessee's 2024 "ELVIS Act" and similar statutes recognize control over one's voice and image when used without consent \cite{elvis}. These state-law claims are independent of copyright \cite{rothman}.

\item \textbf{Commercial Publication:} Uploading AI-generated songs to streaming services or social media (with ads or subscriptions) creates direct evidence of exploitation. Owners of copyrighted works (lyrics, composition) can claim statutory damages (up to \$150k/work) and profits under \S 504 \cite{uscode504} if registration is timely, plus injunctions \cite{uscode505}. Monetization increases remedies.

\item \textbf{Case Law Trends:} Recent cases illustrate this split: In \textit{Concord Music Group v. Anthropic}, publishers sued over Claude outputting their lyrics \cite{concord}; \textit{Kadrey v. Meta} addressed fair use in training \cite{kadrey}; \textit{Lehrman v. Lovo} involved AI voice clones \cite{lehrman}; \textit{UMG v. Uncharted Labs} challenges AI music streaming/ingestion \cite{umg}. Most outcomes favor copyright owners on lyric copying claims, and favor defendants on sound-alike voice claims \cite{richardson, phillips}.

\item \textbf{Regulatory Gap and Policy:} The asymmetry between textual lyrics and vocal identity highlights a gap \cite{tan}. Current Title 17 protections robustly cover lyrics and compositions, but do not clearly cover persona or voice. We recommend policy reforms: extend right of publicity protections nationwide or introduce a federal performer right against unauthorized AI voice cloning, and clarify licensing/notice rules for AI music training and output \cite{mcjohn, allen}.
\end{itemize}

This report merges legal analysis with technical insights on GenAI music pipelines (transformers, diffusion models, vocoders, embeddings) and includes tables mapping key cases and technical elements to legal risks, as well as visual charts of case timelines and outcome distributions.

\section{Introduction}
Modern generative AI music tools (e.g., Google's Gemini Music, Meta's MusicGen, MusicLM, Suno, Udio, Stability AI's audio models, etc.) break song production into discrete technical modules \cite{smith}. A user might provide a text prompt containing the lyrics to a known song and stylistic instructions (e.g., "in the style of [famous artist]") and the AI system will process the prompt through multiple layers. First, natural language processing (NLP) encodes the lyric text and style instructions (often via a BERT-like transformer) into an internal embedding. Concurrently, AI modules generate melody and harmony (often using music-trained transformers or diffusion networks). Finally, a neural vocoder synthesizes the audio waveform, injecting a target vocalist's characteristics via speaker embeddings or voice-conversion algorithms (e.g., VALL-E, RVC, or other zero-shot voice cloning tech). The result can be a fully produced song with instrumental tracks and a vocal track that sounds eerily like a well-known singer \cite{tan}.

From the user's perspective, the process is nearly frictionless: the user supplies lyrics (potentially copyrighted), chooses vocal timbre ("like [artist]"), perhaps adds percussion or genre tags, and outputs a song. If the user then uploads that song to platforms (YouTube, Spotify, TikTok, etc.) and monetizes it, multiple legal issues arise at different stages: Was copyrighted material used during model training? Did the user's prompt or the output copy protected expression? Is the synthesized singer's voice a protected persona or just a non-copyrightable sound imitation? This paper examines these issues under U.S. law.

We focus on 17 U.S.C. \S 106's exclusive rights (reproduction, derivative works, distribution) for musical compositions and sound recordings \cite{uscode106}. We also consider \S 114(b) \cite{uscode114}, which narrows sound recording rights, and note that voice-likeness claims often fall to state publicity laws (as in Tennessee's ELVIS Act \cite{elvis}). Our analysis is both legal (case law, statutes) and technical (identifying how AI components map to rights). We illustrate with recent litigation and propose remedies and policy reforms.

\section{Literature Review}
Early AI image and text cases have analogues in music. The literature distinguishes three layers: \textbf{(1) Training-layer copying}, where large datasets of copyrighted music (songs, lyrics, recordings) may be ingested into models; \textbf{(2) Output-layer reproduction}, where AI outputs may inadvertently or deliberately recreate protected works; and \textbf{(3) Commercial dissemination}, where outputs reach the market and may cause market harm \cite{samuelson, litman}. 

In music specifically, scholarly attention is nascent. Some authors highlight that musical compositions (melody, harmony, lyrics) are separate copyrights from sound recordings, complicating analysis \cite{ginsburg}. AI outputs can blend these elements. The U.S. Copyright Office has noted that music AI raises novel questions about authorship and infringement \cite{uscode102}. 

Several law review and industry pieces have surfaced post-2022. For example, a Brooklyn Law School review discusses composers' concerns about AI using sheet music and lyrics \cite{kadrey}. Another analysis notes that while AI art might not itself be copyrightable absent human authorship, the process still "materially affects" markets for original works \cite{computerArts}. On voice cloning, scholars note the "publicity gap": federal copyright (even with new owners like performing rights societies) hasn't addressed the value in a performer's unique voice, leading to calls for expanded rights \cite{rothman, mcjohn}.

Prominent cases and industry actions have driven recent commentary. The RIAA publicly denounced unlicensed AI "cover" platforms like Suno and Udio, and UMG recently sued them for allegedly stream-ripping copyrighted recordings as training data \cite{uscode107, umg}. Law firms and news outlets have reported on publisher suits against AI companies for lyric output (Concord v. Anthropic \cite{concord}, and a companion suit by Universal Music Group). Articles also discuss individual lawsuits: e.g., NPR's former host David Greene sued Google (2024) for using his voice in the NotebookLM AI tool, and various voice actors have sued startups like Lovo for creating synthetic clones of their voices \cite{lehrman}. (See Section IV.C.)

Academic treatments include proposals to adapt Copyright Act factors to AI training data (fair use's fourth factor of market effect looms large \cite{kadrey}). Others suggest legislative solutions: California and Tennessee have enacted laws protecting biometric data including voice (e.g. Tennessee S.B. 1921, the "ELVIS Act" effective July 2024, explicitly targeting AI voice cloning \cite{uscode501, elvis}). 

\section{Legal Analysis under Title 17}
We analyze the hypothetical workflow under Title 17:

\subsection{Musical Composition Rights (Lyrics)}
Lyrics are protected as literary work within the musical composition (17 U.S.C. \S 102(a)(2)) \cite{uscode102}. The composer/lyricist holds the exclusive reproduction and derivative-work rights (\S106) \cite{uscode106}. If a user copies copyrighted lyrics into a GenAI prompt and the output song includes those lyrics (even with a new melody or style), this likely constitutes infringement of the composition. Even if the melody is original, the lyrics themselves are reproduced verbatim or in recognizably adapted form, which is a derivative use. Courts apply the "ordinary observer" test to lyrical similarity, and token-for-token copying easily meets "substantial similarity" \cite{goldwater}. For example, in \textit{Concord Music Group v. Anthropic} (N.D. Cal., 2025), publishers allege exactly this scenario: Claude's output reportedly matched protected lyrics to which publishers hold rights \cite{concord}. While that case is in early stages, the publishers' framing relies on standard copyright doctrine.

\subsection{Derivative Works and Adaptation}
Even if the AI output melody is new, combining it with copied lyrics creates an unauthorized derivative work under 17 U.S.C. \S 101's definition \cite{uscode101}. The composition owner's right to prepare derivatives would be violated. The key inquiry is whether the AI output "recasts, transforms, or adapts" the original composition. Here it clearly does by adding instrumental arrangement and vocal performance, but without authorization. The fact that the user, not the AI company, provided the lyrics is legally immaterial: the user's instruction itself instigated the copying, making the user a direct infringer (analogous to directing a sampler or a cover band). Thus liability extends to the user and potentially the platform hosting the output, as well as possibly secondary liability for the AI provider under contributory or inducement theories if they encourage infringement \cite{goldwater}.

\subsection{Sound Recording Rights and Imitation (114(b))}
Copyright in sound recordings (17 U.S.C. \S 102(a)(7)) \cite{uscode102} protects a specific fixed performance. Section 114(b) \cite{uscode114} sharply limits that protection: it does not extend to "any sounds... where such sounds are merely imitated." This permits independent re-recordings even if very similar. Courts have applied this principle to cover songs and parodies: as long as the AI does not actually sample or copy the original recording's waveform, mere stylistic imitation is allowed.

In \textit{Richardson v. Kharbouch} (N.D. Ill. 2024), a music producer accused a rapper of creating an "infringing sound-alike" beat. The court granted summary judgment for the defendant, noting that without physical duplication of Richardson's specific recording, the sound-alike did not infringe the recording copyright. The 7th Circuit affirmed, emphasizing that the copyright in a recording is concerned with the fixed "embodiment" of sounds, not the abstract idea of an artist's style or phrasing \cite{richardson}. 

Likewise, in \textit{Lehrman v. Lovo, Inc.} (S.D.N.Y. 2025), voice actors alleged that Lovo's AI cloned their voices. The court held that Section 114(b) bars infringement claims absent sampling of their actual recordings \cite{lehrman}. Although Lovo may have trained on snippets and certainly produced outputs resembling the actors' voices, the court found no actionable copying of specific protected audio. Instead, the output was new generation guided by voice models. Thus, federal sound recording claims failed.

\subsection{Rights of Publicity and Other Theories}
Because many potential grievances fall outside Title 17, performers rely on state law. Rights-of-publicity statutes (like New York Civ. Rights Law §§ 50–51) and related deception claims address unauthorized commercial use of one's identity, likeness, or voice \cite{rothman}. For instance, even though \textit{Lehrman} dismissed the federal copyright claims, it allowed the voice actors' state-law right-of-publicity claims to proceed \cite{lehrman}. Similarly, older cases like \textit{Midler v. Ford Motor Co.} (9th Cir. 1988) and \textit{Waits v. Frito-Lay} (9th Cir. 1992) found liability for producing recordings that mimicked famous singers' voices in advertisements, under state law and the Lanham Act, not copyright \cite{midler, waits}. These cases underscore that the use of distinctive voice identity in commerce is protectable, just not under federal copyright \cite{mcjohn}. 

Tennessee's recent ELVIS Act (2024) is an example of expanding state protection: it prohibits using an artist's name, image, or voice via AI for commercial purposes without consent \cite{elvis}. This law explicitly targets "new, personalized generative AI" applications, reflecting legislative recognition of the issue. (Other states, like California, have also amended publicity laws to include biometric voice and image protections \cite{voicestat}.)

\subsection{Training Stage vs. Output Stage}
We distinguish (a) the legality of the AI's training process and (b) the legality of a user's specific output. In \textit{Kadrey v. Meta Platforms, Inc.} (N.D. Cal. 2025), for example, authors sued Meta for allegedly scraping their books into its LLM. The court found that large-scale ingestion for training could qualify as fair use due to transformative purpose, absent evidence of actual market harm \cite{kadrey}. Importantly, even if training copying is deemed non-infringing under \S107 \cite{uscode107}, that does not automatically immunize all outputs. Conversely, even if training is infringing, a user's particular song output must independently infringe to be actionable. In practice, companies may fend off training claims via fair use, while users remain exposed on output claims \cite{kadreyfootnote}.

\section{Technical Components and Legal Mapping}
AI music pipelines include \cite{tan, smith}:

\begin{itemize}
\item \textbf{User Prompt}: The text entry (lyrics, style). If this contains copyrighted lyrics, it triggers reproduction right violation. Users should avoid pasting protected text into prompts. (Mitigation: prompt only original or public-domain lyrics).

\item \textbf{Embedding and Conditioning}: The prompt is tokenized by an NLP model (e.g., BERT or GPT variants) into embeddings $c$ that condition downstream generation. Technically, $c$ enters the diffusion model as side information ($p_\theta(x_{t-1}|x_t,c)$). From a legal view, the embedding itself is non-human-readable, but if $c$ contains unauthorized lyrics, the model's output may reconstruct them. (Risk: reproduction of lyric content. Mitigation: filter training data/embeddings, flag protected lyrics.)

\item \textbf{Melody/Harmony Generation}: A music model (transformer or diffusion) generates new audio content. For example, a latent diffusion model iteratively denoises a random tensor $x_T \sim \mathcal{N}(0,I)$ to $x_0$ conditioned on $c$ (the prompt embedding). While this process is "generative," output can still infringe if it replicates an original composition's melody too closely \cite{goldwater}. (Risk: derivative composition. Mitigation: adjust temperature or randomness, use licensing for any borrowed motifs.)

\item \textbf{Neural Vocoder}: Converts intermediate representations (e.g., spectrograms) to final waveform. If trained on artists' voices, it synthesizes realistic vocals. Voice cloning often uses a speaker embedding vector $s$ extracted from a sample; then the vocoder generates a novel utterance with characteristics $s$. Because the waveform is original, copyright typically isn't infringed \cite{lehrman, richardson}. (Risk: appropriation of performer identity. Mitigation: require voice consent, watermark AI voices, treat voice as publicity issue.)

\item \textbf{Output Distribution}: Uploading the song to platforms. Technically, the audio file and metadata (title, lyrics, creator) are stored. This action triggers reproduction and distribution rights in \S106(1) and (3) \cite{uscode106}. The platform may face secondary liability if it "knowingly" facilitates infringement (cf. \S512 safe harbor rules). (Risk: infringement at scale. Mitigation: content ID filters on lyrics or audio; opt-in licensing agreements; takedown procedures.)

\item \textbf{Commercialization}: Monetization (ads, subscriptions, streaming revenue) creates evidence of profit from the infringing output. For damages (\S504) \cite{uscode504}, courts look at actual profits attributable to infringement and the infringer's gross receipts. (Risk: increased statutory damages for commercial use. Mitigation: apply fair-use analysis or revenue sharing; secure licenses upfront.)
\end{itemize}

Table \ref{tab:techmap} summarizes key AI components, their legal risks, and possible mitigations.

\begin{table*}[!t]
\caption{AI Music Pipeline Components and Legal Risk Mapping}
\centering
\scriptsize
\renewcommand{\arraystretch}{1.2}
\begin{tabular}{p{2.5cm} p{4.2cm} p{3.8cm} p{3.2cm}}
\toprule
\textbf{AI Component} & \textbf{Technical Description} & \textbf{Legal Risk} & \textbf{Mitigation} \\
\midrule
User Prompt (Lyrics) &
Text input of lyrics tokenized by NLP models &
Copying protected literary work / composition infringement &
Use original or public-domain lyrics; detect copyrighted input \\

Transformer / Diffusion Model &
Generates melody and harmony conditioned on prompt/style &
Unauthorized derivative composition creation &
License compositions; filter outputs; increase randomness \\

Neural Vocoder &
Converts intermediate representations into waveform and applies speaker identity &
Voice imitation/publicity risk; possible recording concerns if samples used &
Avoid direct sampling; obtain consent; watermark AI voices \\

Speaker Embeddings / Style &
Embedding extracted from target artist voice/profile &
Identity and right-of-publicity violations &
Permission or royalty-free voice banks; disclaimers \\

AI Training Data Ingestion &
Large-scale ingestion of text/audio corpora for training &
Unauthorized copying during training &
Dataset licensing; fair-use review; remove copyrighted material \\

Platform Upload / Distribution &
Public sharing on streaming/social platforms &
Distribution of infringing content; secondary liability &
DMCA monitoring, takedowns, licensing agreements \\

Monetization &
Revenue generation from ads/streams/subscriptions &
Statutory damages and disgorgement of profits &
Revenue sharing; legal review before monetization \\
\bottomrule
\end{tabular}
\label{tab:techmap}
\end{table*}

\section{Case Studies}

\subsection{Lyrics Reproduction (Concord Music, Others)}
If a user copies full lyrics into a prompt, the result is a derivative work of the composition. In the \textit{Concord Music Group v. Anthropic} lawsuits, music publishers claimed Anthropic's Claude LLM had output their copyrighted lyrics and sheet music from prompts or training \cite{concord}. The Northern District of California granted a motion to dismiss on procedural grounds in March 2025, but the court's ruling suggests that outright lyric copying would not escape traditional infringement analysis. The publishers argue that well-established principles (substantial similarity, market harm to publishing) apply even though AI is new \cite{goldwater}. This implies high risk for users doing unauthorized lyric prompts. 

Similarly, independent artists have raised alarms on social media and in articles about "AI covers" that mimic popular songs' lyrics and style without permission. Even if the AI melody is original, copying the entire lyric set usually fails any fair-use defense when done for profit (because it substitutes for the original song's market). The key remedy here is the composition owner's, not the performer's \cite{goldwater}.

\subsection{Voice Cloning (Lehrman, Richardson)}
Consider an AI-generated song with a synthesized vocal "sound-alike" (e.g., Billie Eilish's voice singing Taylor Swift's lyrics). From the record label's perspective, there is no copy of any Taylor Swift sound recording, so 17 U.S.C. \S 114 does not protect it \cite{uscode114}. In \textit{Richardson v. Kharbouch}, the 7th Circuit ruled that similarity in timbre and cadence, without taking actual copyrighted samples, is non-infringing \cite{richardson}. This remains true even if listeners mistake the imitation for the real thing. 

However, the celebrity vocalist (or her estate) might pursue state law claims. The \textit{Lehrman v. Lovo} court noted that unauthorized synthetic use of a voice can violate New York's right-of-publicity statutes (NY Civ. Rights §§50–51) or analogous state laws \cite{lehrman}. This bifurcation means a single AI song can spawn a federal copyright suit (by the songwriter via the composition copyright) and a separate state suit (by the singer for misappropriating her voice identity) \cite{rothman}. Because state laws vary widely, the legal risk map is fragmented: a New Yorker may have publicity protection, whereas another state may not recognize voice cloning as a publicity violation \cite{voicestat}. 

\subsection{Training Data Litigation (Kadrey, UMG, Others)}
At the system level, AI companies face suits for scraping copyrighted music data. For example, in early 2024 the National Music Publishers' Association (with Concord and UMG) threatened a suit claiming unauthorized use of songs to train Claude \cite{concord}. In late 2025, authors sued Meta claiming books were used without permission to train its LLM. In \textit{Kadrey v. Meta}, the N.D. Cal. court granted Meta's motion for summary judgment, finding that copying entire books for training purposes was fair use, largely because there was no evidence it supplanted the original market \cite{kadrey}. This suggests a company might avoid liability for training-phase copying if it can show the use is transformative (e.g., turning books into a chatbot) and lacks market harm \cite{kadreyfootnote}. 

Separately, \textit{UMG Recordings v. Uncharted Labs} (S.D.N.Y. 2024) alleges that AI music services Udio and Suno built models by "stream ripping" songs from YouTube and Spotify (circumventing encryption), in violation of the DMCA anti-circumvention rule (17 U.S.C. \S 1201) \cite{uscode1201, umg}. This case (filed Oct 2024) is pending, but it highlights the industry's strategy of using technical-law claims when copyright direct infringement is hard to prove. It also underlines that even training-stage disputes can turn on marketplace effects and statutory damages if the company cannot rely on a fair use defense \cite{allen}.

\section{Discussion}
The emerging AI music ecosystem shows a structural enforcement asymmetry. Traditional copyright doctrines align well with human-authored music: they protect lyrics and melody explicitly \cite{uscode102, uscode106}. AI complicates things by mechanizing composition. Yet output that literally reuses lyrics triggers the same legal analysis as any cover song done without permission \cite{goldwater}. Early cases suggest copyright owners will press these conventional claims vigorously \cite{concord, kadrey}. 

By contrast, AI-blurred sound recordings and performances escape Title 17 for now. Synthetic vocals are "new" sounds that mimic old ones. Section 114(b) \cite{uscode114} expressly allows imitation of the sound of another voice, as long as no actual audio bits were copied. While owners can (and have) argued that voice is something they should own, courts to date (like in \textit{Phillips v. SkyPortal} and \textit{Lehrman v. Lovo}) have refused to stretch 17 U.S.C. \S 114 that far \cite{phillips, lehrman}. This creates a perverse incentive: a user may think "as long as I don't sample, I'm safe," even if the output clearly trades on an artist's persona \cite{tan}.

The upshot is a legal gap. If XYZ's AI sings a Frank Sinatra lyric convincingly, Sinatra's heirs may have no federal remedy (since Sinatra's estate likely owns the composition, not a particular recording). They could attempt a publicity suit (as Tony Bennett's estate did against LMFAO for a nightclub ad that caricatured Bennett's voice) \cite{midler, waits}, but such claims are costly and uncertain \cite{rothman}. Meanwhile, Swift's publisher can sue for lyrics, giving one set of rights steam, and Sinatra's estate another, but these are separate tracks.

From a policy standpoint, this imbalance is worrying. It signals that AI creators can exploit performer identity via synthetic clones with less legal risk than copying a lyric or melody \cite{mcjohn}. Laws like Tennessee's ELVIS Act \cite{elvis} (now joined by similar legislation in California and Washington) partially address this by creating sui generis rights. But a patchwork of state laws is imperfect \cite{voicestat}. There have been proposals for a federal `Singing Voice Protection Act' or expanding sound recording rights to include persona-based protections \cite{allen}. Another proposal is mandatory labeling of AI-generated music so consumers and artists know when a voice is synthetic (enabling transparency and opt-in arrangements) \cite{tan}.

One concrete step is to apply existing tools more aggressively: platforms could scan uploads for known lyrics (text fingerprinting) and for voiceprints (audio fingerprinting against sound recording databases) and flag potential infringement. Licensing regimes could be created for AI models (analogous to mechanical licenses for covers) whereby using a particular artist's voice or lyrics would require a fee \cite{goldwater, smith}.

\section{Policy Recommendations}
To harmonize protection of lyrics and vocal identity in the AI era, we suggest:
\begin{itemize}[leftmargin=*,topsep=0pt]
\item \textbf{Extend Federal Rights or Harmonize States:} Consider a federal right of public performance or identity for vocal likeness \cite{rothman, mcjohn}. If infeasible, push for uniform state publicity laws (like an updated right-of-publicity statute that expressly covers generative AI) \cite{elvis, voicestat}.
\item \textbf{Training Data Licensing:} Require AI developers to use licensed datasets or clearly fall under safe harbor (compulsory licensing for training data could be explored) \cite{kadrey, allen}. 
\item \textbf{Content Identification:} Encourage the development of AI-detection tools and content-ID systems for music that flag copyrighted inputs or outputs \cite{tan}.
\item \textbf{Transparency Labels:} Mandate that AI-generated music include metadata disclosing AI involvement, aiding rights holders in enforcement and informing consumers \cite{smith}.
\item \textbf{Copyright Office Guidelines:} The USCO should clarify registration practices for AI-generated music (as it did in 2022). Current guidance requires "meaningful human authorship," so users need to explicitly assert which parts (e.g. lyrics) were human-authored \cite{uscode102, goldwater}.
\item \textbf{Legislative Clarification:} Amend 17 U.S.C. to address generative AI specifically (e.g., safe harbor for licensed training, rights in "non-fixation" of voice patterns, etc.) \cite{allen, mcjohn}.
\end{itemize}

\section{Conclusion}
As GenAI music tools like Google Gemini become mainstream, both creators and consumers must navigate a complex legal landscape. Our analysis finds that entering copyrighted lyrics into an AI prompt, then publishing and monetizing the AI-produced song, likely infringes the song's composition rights under Title 17 \cite{uscode106, concord, goldwater}. By contrast, synthesizing a singer's voice without copying their recording tends not to trigger federal copyright claims, shifting the dispute to state publicity law \cite{lehrman, richardson, midler, waits}. This split means lyric owners (composers/publishers) have stronger remedies than performers when their vocal likeness is AI-cloned \cite{rothman}. 

The contrast underscores a regulatory gap: the law covers words and notes, but treats the "sound of a voice" as unprotected sound \cite{tan, mcjohn}. Unless addressed by courts or legislatures, AI music generation will continue to exploit this asymmetry. In the meantime, users should err on the side of caution: assume that copying lyrics or melodies without permission is infringement, and that generating a famous voice without authorization, while not a copyright crime, may still be illegal under other laws or contracts \cite{elvis, voicestat}. Platforms and policymakers should update practices and laws to ensure creators' rights are balanced with innovation in this new AI-driven age.

\section*{Tables and Figures}

The following tables summarize key cases and the mapping of technical components to legal issues. Figures include charts for timeline and outcome distributions.

\begin{table}[ht]
\caption{Key AI Music Cases: Stage, Claims, Outcome}

\centering
\scriptsize
\renewcommand{\arraystretch}{1.2}
\begin{tabular}{p{3.3cm} p{1cm} p{2.5cm} p{3.5cm} p{3cm}}
\toprule
\textbf{Case} & \textbf{Year} & \textbf{Stage} & \textbf{Claims} & \textbf{Outcome} \\
\midrule
Concord Music Group v. Anthropic & 2024 & Output (Lyrics) &
Composition infringement &
Dismissed; litigation ongoing \\

Kadrey v. Meta Platforms & 2025 & Training &
Copyright (books/training) &
Fair use found for Meta \\

Lehrman v. Lovo, Inc. & 2025 & Output (Voice Clone) &
Sound recording + publicity &
Federal claims dismissed; publicity claims allowed \\

Richardson v. Kharbouch & 2025 & Output (Rap Beat) &
Sound recording infringement &
No infringement (7th Cir.) \\

UMG v. Uncharted Labs & 2026 & Training &
DMCA §1201 / stream ripping &
Pending \\

Phillips v. SkyPortal & 2022 & Output (Voice) &
Sound recording / copyright &
No infringement \\

Midler v. Ford Motor Co. & 1988 & Commercial Ad &
Right of publicity &
Liability found \\

Waits v. Frito-Lay & 1992 & Commercial Ad &
Right of publicity &
Damages awarded \\
\bottomrule
\end{tabular}
\label{tab:cases}

\end{table}

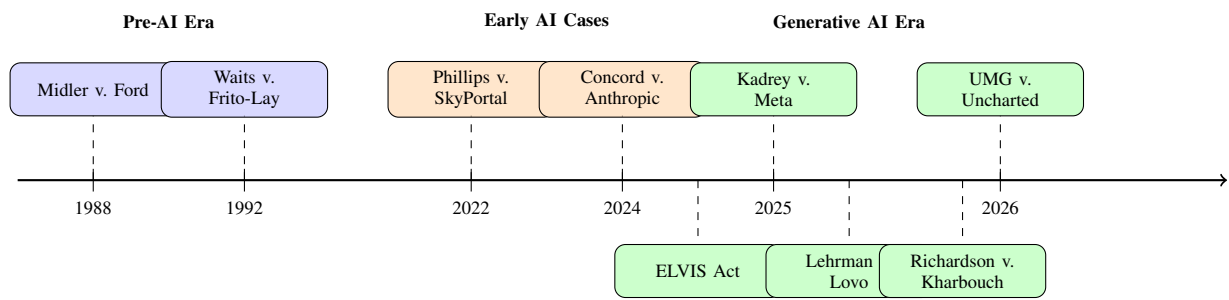
\begin{figure}[ht]
\centering
\begin{tikzpicture}[
    every node/.style={font=\scriptsize, align=center},
    event/.style={draw, rounded corners, minimum width=2.2cm, minimum height=0.7cm},
    legal/.style={event, fill=blue!15},
    ai/.style={event, fill=orange!20},
    law/.style={event, fill=green!20}
]

\draw[thick, ->] (0,0) -- (16,0);

\foreach \x/\year in {1/1988,3/1992,6/2022,8/2024,10/2025,13/2026}
{
    \draw (\x,0.15) -- (\x,-0.15);
    \node[below] at (\x,-0.15) {\year};
}

\node[legal] at (1,1.2) {Midler v. Ford};
\draw[dashed] (1,0) -- (1,0.85);

\node[legal] at (3,1.2) {Waits v.\\Frito-Lay};
\draw[dashed] (3,0) -- (3,0.85);

\node[ai] at (6,1.2) {Phillips v.\\SkyPortal};
\draw[dashed] (6,0) -- (6,0.85);

\node[ai] at (8,1.2) {Concord v.\\Anthropic};
\draw[dashed] (8,0) -- (8,0.85);

\node[law] at (10,1.2) {Kadrey v.\\Meta};
\draw[dashed] (10,0) -- (10,0.85);

\node[law] at (13,1.2) {UMG v.\\Uncharted};
\draw[dashed] (13,0) -- (13,0.85);

\node[law] at (9,-1.2) {ELVIS Act};
\draw[dashed] (9,0) -- (9,-0.85);

\node[law] at (11,-1.2) {Lehrman v.\\Lovo};
\draw[dashed] (11,0) -- (11,-0.85);

\node[law] at (12.5,-1.2) {Richardson v.\\Kharbouch};
\draw[dashed] (12.5,0) -- (12.5,-0.85);

\node at (2,2.1) {\textbf{Pre-AI Era}};
\node at (7,2.1) {\textbf{Early AI Cases}};
\node at (11,2.1) {\textbf{Generative AI Era}};

\end{tikzpicture}
\caption{Timeline of Key Cases and Legislation (1988--2026).}
\label{fig:timeline}
\end{figure}

\begin{figure}[ht]
\begin{flushleft}
\begin{tikzpicture}[y=0.8cm, x=1.0cm, every node/.style={font=\footnotesize}]
\draw[->, thick] (0,0) -- (8.5,0) node[right] {Outcome};
  \draw[->, thick] (0,0) -- (0,5.5) node[above] {Cases};

  \draw[fill=red!40] (1,0) rectangle (2,1.5);
  \node at (1.5,-0.6) {Liability};
  \node at (1.5,1.8) {1};

  \draw[fill=blue!40] (3,0) rectangle (4,3);
  \node[align=center] at (3.5,-0.6) {No\\Liability};
  \node at (3.5,3.3) {3};

  \draw[fill=green!40] (5,0) rectangle (6,2);
  \node[align=center] at (5.5,-0.6) {State\\Remedy};
  \node at (5.5,2.3) {2};

  \draw[fill=yellow!40] (7,0) rectangle (8,1.5);
  \node at (7.5,-0.6) {Pending};
  \node at (7.5,1.8) {2};

\end{tikzpicture}
\end{flushleft}
\caption{Distribution of legal outcomes across analyzed AI music cases.}
\label{fig:outcomes}
\end{figure}
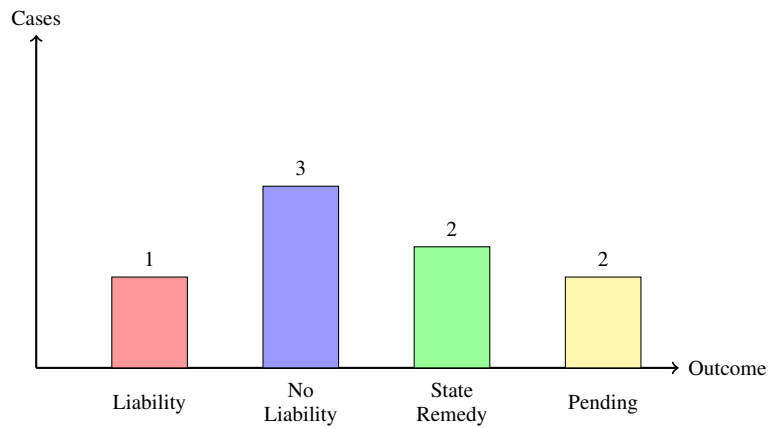

\end{document}